\documentclass[aps,prl,showpacs,twocolumn,citeautoscript,superscriptaddress]{revtex4-1}

\usepackage{graphicx}
\usepackage[bookmarks=true]{hyperref}
\newcommand{\vect}[1]{{\vec{#1}}}
\newcommand{\dens}{n} %Changeable eln density - n or rho
\newcommand{ \cro}{\mbox{Cr$_2$O$_{3}$}}
\newcommand{ \lnp}{\mbox{LiNiPO$_{4}$}}

\begin{document}

\title{The importance of the electronic contribution to linear magnetoelectricity}

\author{Kris~T.~Delaney}
\thanks{Authors contributed equally.}
\affiliation{Materials Department, University of California, Santa Barbara, CA 93106-5050, USA}
\author{Eric~Bousquet}
\thanks{Authors contributed equally.}
\affiliation{Materials Department, University of California, Santa Barbara, CA 93106-5050, USA}
\affiliation{Institut de physique (B5), Universit\'e de Li\`ege, B-4000 Sart Tilman, Belgium}
\author{Nicola~A.~Spaldin}
\affiliation{Materials Department, University of California, Santa Barbara, CA 93106-5050, USA}

\date{\today}

\begin{abstract}
We demonstrate that the electronic contribution to the linear
magnetoelectric response, usually omitted in first-principles studies, can be
comparable in magnitude to that mediated by lattice distortions, even for materials in
which responses are strong.
Using a self-consistent Zeeman response to an applied magnetic field 
for noncollinear electron spins, we show how electric polarization emerges in linear
magnetoelectrics through both electronic- and lattice-mediated components --
in analogy with the high- and low-frequency dielectric response to an electric field.
The approach we use is conceptually and computationally simple, and can be
applied to study both linear and non-linear responses to magnetic fields.
\end{abstract}


\maketitle

Linear magnetoelectrics are materials which respond with a change in electric polarization to
a magnetic field, or conversely with a change in magnetization to an electric field:
$P_i = \alpha_{ij}H_j; \quad M_j = \alpha_{ij}E_i$, 
where $\alpha$ is the linear magnetoelectric tensor.

The research challenges in identifying materials with useful linear magnetoelectric (ME) responses
are threefold: (i)  Symmetry requirements that both space-inversion and time-reversal symmetries be broken
are satisfied by few materials, (ii)  materials satisfying these criteria tend to do so only in
phases that develop at relatively low temperatures, and (iii)  most of the MEs discovered to date
have weak responses.
Recently, a number of developments have led to a significant revival of activity in
the search for novel magnetoelectric materials, including the observation that multiferroics
can have strong ME responses\cite{Fiebig:2005}, and the identification 
of the microscopic couplings that are responsible for strong and weak ME 
responses\cite{Kimura_et_al:2003,Goto_et_al:2004,Katsura/Nagaosa/Balatsky:2005,Picozzi_et_al:2007,Delaney/Mostovoy/Spaldin:2009}.

First-principles methods are emerging\cite{Iniguez:2008,Delaney/Mostovoy/Spaldin:2009} 
as a valuable tool for computing the strength of
$\alpha$ in
real materials without any empirical input. 
The methods are becoming sufficiently reliable to be used in a predictive capacity in searching for new
ME materials.
While many approaches have been explored or can be envisaged for computing $\alpha$, ranging from a self-consistently applied
electric or magnetic field, to quantum-mechanical perturbation theory in the applied field, by
far the most successful and widely used approach for applications to date has been a 
linear-response approach based on the lattice-dynamical quantities\cite{Iniguez:2008,Delaney/Mostovoy/Spaldin:2009,Wojdel/Iniguez:2009}.
However, this approach computes only the so-called ``lattice-mediated'' part of $\alpha$ and ignores
purely electronic contributions to the response. 
The common justification is that such contributions are expected to be weak, just as in strong dielectrics
the electronic response is negligible compared to the ionic contribution.
In this Letter, we demonstrate, using an alternative numerical approach involving a self-consistently 
applied magnetic field, that the purely electronic magnetoelectric response can in fact be large, even
in materials with relatively strong $\alpha$ in which one might expect the response to be dominated by 
lattice mechanisms.
In fact, the electronic and ionic contributions can be of similar magnitude and opposite sign, which
can lead to a weak total response while lattice-only methods would predict it to be large.

We begin by defining the different contributions to $\alpha$. 
We define a ``clamped-ion'' contribution that accounts for
magnetoelectric effects occuring in an applied field with all ionic degrees of freedom remaining frozen.
This purely electronic contribution, $\alpha^\mathrm{el}$, is the response that would be
measured for high-frequency fields, in analogy with the static-high-frequency
dielectric response in insulators, $\epsilon_\infty$. 
The remaining part of the response, which emerges at low frequency so that the ions
(and in principle the lattice parameters\cite{Wojdel/Iniguez:2010aXv})
respond to the field, we label $\alpha^\mathrm{latt}$.
$\alpha^\mathrm{latt}$ is the difference between the total response and the electronic, so that
$\alpha^\mathrm{tot}=\alpha^\mathrm{el}+\alpha^\mathrm{latt}$, and it is the part of the
response that is accessible from linear-response theory using lattice-dynamical quantities\cite{Iniguez:2008}.
(Note that $\alpha^\mathrm{latt}$ also includes some electronic response through the dynamical charges).

Evaluating $\alpha^\mathrm{el}$ requires a full quantum-mechanical treatment of the response, either
using perturbation theory or a self-consistently-applied finite field. 
In this work, we use an applied magnetic field to compute $\alpha$. 
Due to difficulties with periodic boundary conditions when using a full vector potential 
in the electronic Hamiltonian, we restrict our field to act only on the electron spin as a 
self-consistent Zeeman field. 
This means that we omit contributions to $\alpha$ that are derived from the orbital magnetic
response\cite{Malashevich/Souza/Coh/Vanderbilt:2010aXv}, which are expected to be significantly weaker than the spin-derived response
for most systems.

\paragraph*{Computational Approach:}
Calculations of ME responses require  a treatment of noncollinear spin orders.
Within the Kohn-Sham framework, noncollinearity is handled by generalizing the 
orbitals to be complex spinors, resulting in a $2\times2$ spin-density matrix 
($n_{\sigma\sigma^\prime}$) that
allows the magnetization density to vary in both magnitude and direction throughout
the system.

In order to apply a magnetic field, we begin by making a Legendre transform of the noncollinear-spin 
Hohenberg-Kohn energy functional:
\begin{equation}
\Lambda\left[n_{\sigma\sigma^\prime}\left(\vect{r}\right); \vect{H}\right] = E_\mathrm{HK}\left[n_{\sigma\sigma^\prime}\left(\vect{r}\right)\right] - \mu_0\vect{H}.\vect{\mu}_\mathrm{tot},
\end{equation}
where $E_\mathrm{HK}$ is 
the usual zero-field energy functional (consisting of non-interacting 
kinetic energy, external electrostatic energy, Hartree and exchange-correlation terms, and
the Madelung energy of ion-ion interactions). $\vect{H}$ is the auxiliary magnetic field
applied to spin degrees of freedom and $\vect{\mu}_\mathrm{tot}$ is the total spin magnetic moment of
the system (including the Land\'e $g$ factor).
The spatially varying magnetic moment, $\vect{\mu}\left(\vect{r}\right)$, can be found from the 
four components of $\dens_{\sigma\sigma^\prime}$, and $\vect{\mu}_\mathrm{tot}$ is its spatial integral.
Subsequently, we variationally minimize $\Lambda$ with respect to
single-particle Kohn-Sham spinor orbitals, subject to the usual constraints on 
orthonormality and conservation of total particle number. 
The result is a term in the Kohn-Sham potential, acting on the spinor orbitals, that
imposes the external magnetic field on the noncollinear spin density.
Practically, this term simply shifts the relative external potential for
each of the four spin manifolds:
\begin{equation}
\Delta V_{\sigma \sigma^\prime} = -\frac{g}{2}\mu_B\mu_0\left(
\begin{array}{cc}
H_z & H_x+iH_y \\
H_x-iH_y & -H_z \end{array}
\right),
\label{eqn:constrpot_spinor}
\end{equation}
which is trivially compatible with periodic boundary conditions, and which clearly reduces to 
the collinear case, with the field providing
a different Fermi level for ``up'' and ``down'' spin channels, if $\vect{H} = \left(0,0,H_z\right)$.

We implement Eq.~\ref{eqn:constrpot_spinor} into the Vienna Abinitio Simulation Package (VASP)\cite{Kresse/Furthmuller:1996}.
We employ a plane-wave basis set for expanding the electronic wave functions and density, and 
PAW potentials\cite{Kresse/Joubert:1999} are used for core-valence separation. 
We note that it is important to disable symmetrization of the wave functions in the Brillouin
zone since application of a magnetic field leads to an electronic
structure that breaks the crystal symmetry. 
We self-consistently include spin-orbit coupling in all of our calculations.
This is required to obtain a magnetoelectric response for those materials in which
the spin-lattice coupling derives from relativistic effects such as the antisymmetric ($\vect{s}_i\times\vect{s}_j$)
Dzyaloshinski\u{\i}-Moriya interaction.
All of our calculations simulate a single magnetic domain, which corresponds to
experiments in which a poling procedure has been performed.
Since different antiferromagnetic domains contribute to $\alpha$ with different signs,
measurements otherwise tend to provide a lower bound on the single-domain response.

\paragraph*{Transverse magnetoelectric response of Cr$_2$O$_3$:}
\cro{}, the first ME material to be discovered, remains the best-studied and prototypical linear
magnetoelectric.
Cr$_2$O$_3$ adopts the space group R$\bar{3}$c in the ground state, with Cr and O occupying
Wyckoff positions 4c and 6e in the rhombohedral setting.
We work with the experimental 
volume ($95.9$\AA{}$^3$) and rhombohedral angle ($55.13^\circ$)\cite{Finger/Hazen:1980}, but
fully optimize the ion coordinates to the LDA+U ground-state (Cr: $x=0.1536$; O: $x=0.9426$).
For the exchange-correlation potential, we use the Dudarev form of LSDA+$U$ with $U_\mathrm{eff}=2.0$\,eV\cite{Iniguez:2008}.
In the ground-state G-type antiferromagnetic (AFM) ordering, the free energy of the system contains
two symmetry allowed linear magnetoelectric couplings and is of the form\cite{Dzyaloshinskii:1959}
\begin{equation}
F = -\alpha_\perp\left(E_xH_x + E_yH_y\right)-\alpha_\parallel E_zH_z.
\end{equation}
We note that the response parallel to the trigonal (easy) axis, $\alpha_\parallel$, has been
experimentally demonstrated to be close to zero at zero temperature\cite{Folen_et_al:1961}.
This small low-temperature response is expected: In the absence of quantum fluctuations and
orbital contributions, the low-temperature ME response should tend to zero because
the spin-only parallel \emph{magnetic susceptibility} vanishes at $T=0$\,K.
Our calculations, which are formally at zero temperature and do not include
quantum spin fluctuations or the orbital magnetic response, do give $\alpha_\parallel=0$ as expected.
The temperature dependence of $\alpha_\parallel$ has recently been explained using an alternative
first-principles scheme\cite{Mostovoy/Scaramucci/Delaney/Spaldin:2010aXv}.

\begin{figure}[h!]
\begin{center}
\resizebox{0.85\columnwidth}{!}{\includegraphics[angle=0]{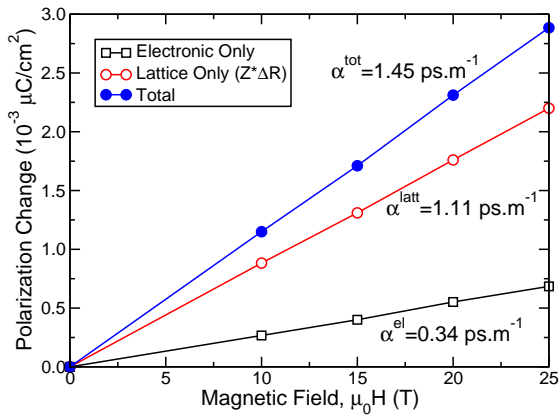}}
\end{center}
\caption{(Color online) The separate contributions to the transverse magnetoelectric
response of Cr$_2$O$_3$ calculated using Eq.~\ref{eqn:constrpot_spinor}. 
The clamped-ion response, $\alpha^\mathrm{el}$, is shown as open squares, and contributes
approximately one fourth of the total response (filled circles).  
The remainder of the response is due to structural distortions in the applied field
(open circles), computed using the Born effective charge tensor.
}
\label{fig:cr2o3}
\end{figure}

For the transverse response of \cro{}, $\alpha_\perp$, we first compute the
electronic ME contribution with clamped ions.
Upon application of a Zeeman field perpendicular to the collinear spin axis, the Cr spins cant so
that the unit cell acquires a net magnetization.
Our calculated spin-only magnetic susceptibility is $\chi^M_\perp = 1.9\times10^{-3}$ in dimensionless
SI units, which compares favorably to the experimental value\cite{McGuire_et_al:1956} of 
$1.7\times10^{-3}$ at $78$\,K.
With the new spin configuration generated by the applied magnetic field, spin-orbit coupling leads
to an electric polarization. 
Note at this stage that the lattice has not been permitted to relax in response to the field.
We compute the electric polarization using the Berry phase approach\cite{King-Smith/Vanderbilt:1993}.
The results are shown as open squares in Fig.~\ref{fig:cr2o3}, 
with our calculated $\alpha^\mathrm{el}_\perp = 0.34$\,ps.m$^{-1}$.

The canting of spins in the applied magnetic field also causes spin-orbit-driven forces on ions.
The induced forces are very small, mirroring the observed weakness of the linear ME effect in \cro{}, so
high numerical quality must be achieved in studying the structural distortion. 
We carefully converge our ionic forces with respect to basis-set size ($E_\mathrm{cut}=700$\,eV),
$k$-point sampling ($6\times 6\times 6$), and self-consistent field iteration.
We then choose a small tolerance to which forces are eliminated ($< 5$\,$\mu$eV \AA{}$^{-1}$).

Once the structural distortions
are found for a given magnetic-field strength, we quickly estimate the ionic polarization by
multiplying the ion displacements by the full Born-effective-charge tensor (see open circles of Fig.~\ref{fig:cr2o3}).
This procedure leads to the lattice-mediated ME response, $\alpha^\mathrm{latt}_\perp=1.11$\,ps.m$^{-1}$. 
We also computed the same quantity using our parameters with the lattice-dynamical method 
for $\alpha^\mathrm{latt}$ introduced previously\cite{Iniguez:2008} and find $0.9$\,ps.m$^{-1}$, 
in reasonably good agreement with the applied magnetic field approach.

With the lattice distortion included for each magnetic-field strength, we now fully recompute the
electric polarization using the Berry-phase approach \emph{with the magnetic field applied}.
This procedure yields the full spin-mediated $\alpha$, both electronic and lattice contributions,
and is shown as filled circles in Fig.~\ref{fig:cr2o3}.
We find $\alpha_\perp=1.45$\,ps.m$^{-1}$ ($=4.3\times10^{-4}$\,emu-CGS), 
in excellent agreement with summing $\alpha^\mathrm{el}_\perp$ and $\alpha^\mathrm{latt}_\perp$ as expected. 
From this analysis, it is clear that the electronic response, $\alpha^\mathrm{el}$, contributes one fourth
of the total spin-driven magnetoelectric response of \cro{}.
The total response that we compute is in good agreement with zero-temperature extrapolations of 
experimental measurements, which range\cite{Wiegelmann_et_al:1994,Kita/Siratori/Tasaki:1979,Iniguez:2008} from $2$---$4.7\times10^{-4}$\,emu-CGS.

\paragraph*{Off-diagonal ME response of LiNiPO$_4$:}
The lithium orthophosphates LiMPO$_4$ (M=transition metals) are currently
attracting much interest because of their potential use as cathodes for
Li-ion batteries, as well as for their large magnetoelectric responses and the recent observation
of ferrotoroidic domains in LiCoPO$_4$\cite{vanaken2007}.

Here we focus on \lnp{} as a test case for the size of the electronic ME response. 
\lnp{} has space group Pnma with $28$ atoms in the primitive unit cell and
four Ni$^{2+}$ magnetic sites.
We again use the experimental\cite{Abrahams/Easson:1993} lattice parameters for our
calculations ($a=10.032$\,\AA{}, $b=5.854$\,\AA{}, $c=4.677$\,\AA{})
while the ionic coordinates are fully relaxed to the LDA+U ground state.

We first address the zero-field magnetic structure.
Experimentally\cite{Jensen2009}, the magnetic structure has been characterized as C-type AFM
with spins oriented predominantly along the $c$ axis, combined with a weak spin canting of A-type AFM order
along the $a$ axis. 
The reported zero-field canting angle, $\theta$, in the Ni compound is $7.8^\circ$.

With a plane-wave cutoff of $500$\,eV and $4\times4\times2$ $k$-point sampling,
we qualitatively reproduce the observed magnetic structure in our calculations.
However, the precise canting angle is strongly sensitive to the \emph{intra}-atomic $J$ parameter
used in the Liechtenstein LDA+$U$ procedure.
For $U=5$\,eV, a common value for Ni$^{2+}$, the canting angle $\theta$ varies from $1.6^\circ$ to
$7.6^\circ$ as $J$ is modified from $0.0$ to $1.7$\,eV.
Hence, a relatively large $J$ appears to be important for quantitative agreement
with the magnitude of the reported spin canting.
The observed sensitivity arises from the double-counting term in the LSDA+U potential, in which
the off-diagonal elements (acting on the non-collinear part of the spin density matrix)
are determined by $J$ alone.
The canting angle is quite insensitive to our other simulation parameters,
including the on-site Coulomb interaction, $U$.

\begin{figure}[ht]
\begin{center}
\includegraphics[width=1.00\columnwidth,keepaspectratio=true]{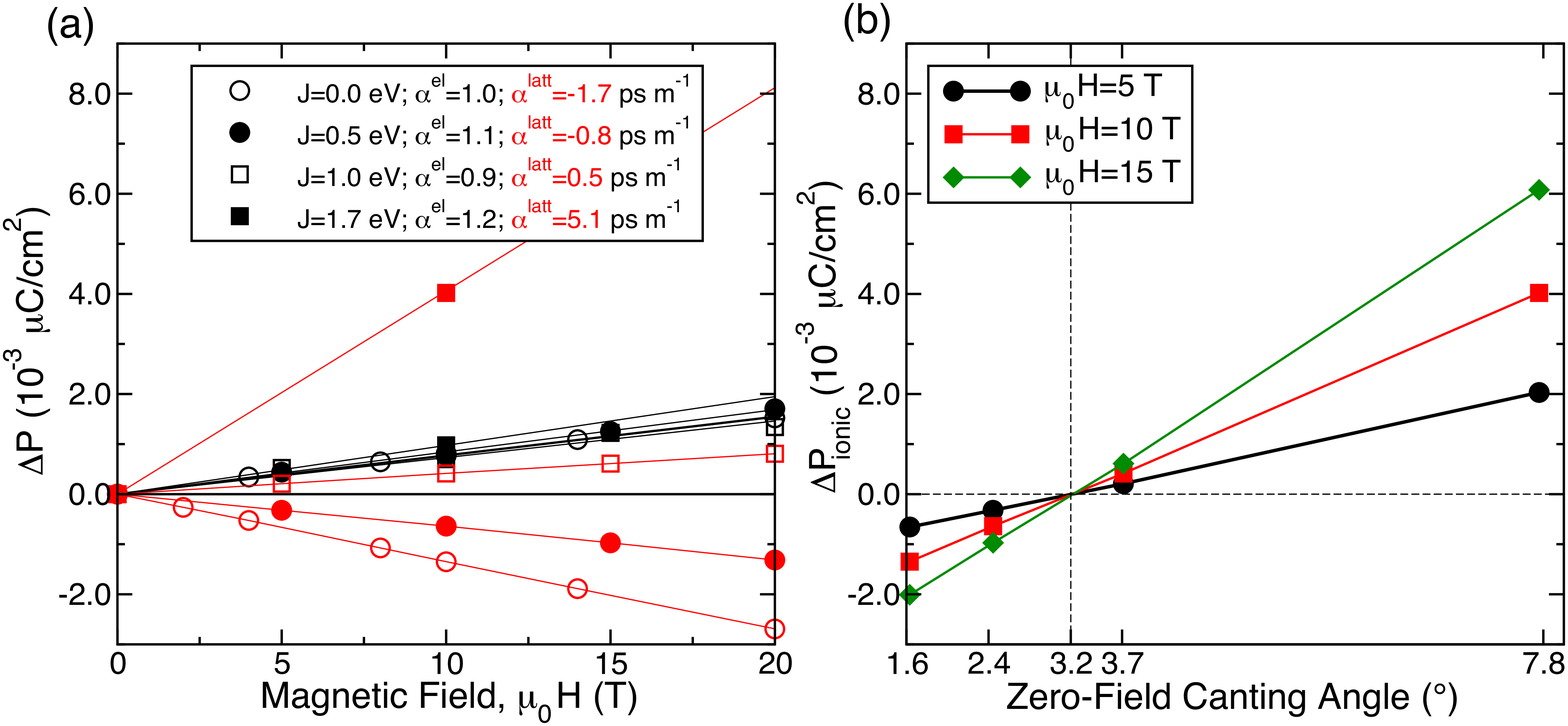}
\caption{(Color online) LiNiPO$_4$ data under magnetic field $\vect{H}\parallel a$: 
(a) Clamped-ion (black) and ionic (red) contributions to the polarization versus the magnetic field for different values of the intra-atomic Hund's coupling $J$. 
(b) Ionic polarization versus canting angle for three values of the magnetic field. 
}
\label{fig:lnp}
\end{center}
\end{figure}

\begin{figure}[ht]
\begin{center}
\includegraphics[width=0.90\columnwidth,keepaspectratio=true]{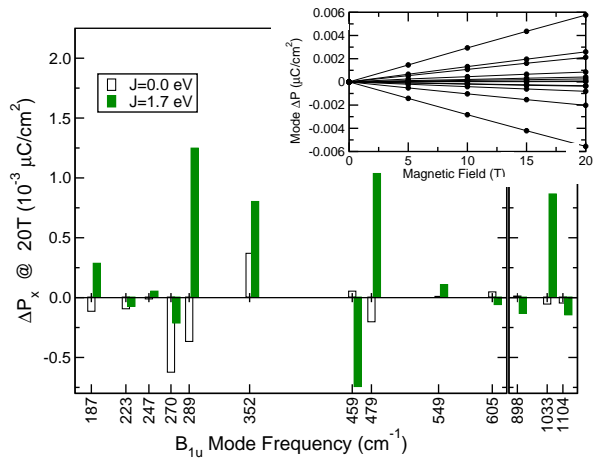}
\caption{(Color online) Mode contributions to the 
  ionic polarization of LiNiPO$_4$ for a field strength of $20$\,T.
  Inset: linear evolution of each of the 13 $B_{1u}$ (polar) lattice modes with applied
  field.
}
\label{fig:lnpb1u}
\end{center}
\end{figure}

We now turn to the magnetoelectric response of \lnp{}.
For the magnetic point group $mm^\prime m$, the magnetoelectric tensor admits only two
non-zero components: $\alpha_{xz}$ and $\alpha_{zx}$.
We focus on the larger component $\alpha_{zx}$, which we access by computing
the Berry-phase polarization along $c$ while
applying a magnetic field along the $a$ axis ($\vect{H}$ perpendicular to the easy axis).
In Fig.~\ref{fig:lnp}a we report both the clamped-ion and lattice-mediated 
magnetoelectric response under this orientation of magnetic field. 
For each of the four different values of the intra-atomic $J$, we 
find $\alpha^\mathrm{el}_{zx}\sim1.0$\,ps.m$^{-1}$.
However, as might be expected from the $J$ dependence of the spin canting angle,
the ionic contributions to $\alpha$ is strongly sensitive to $J$.
For $J=0.0$\,eV, $\alpha^\mathrm{latt}_{zx}$ is moderately large and negative. With increasing
$J$, this component of the response progressively increases, eventually turning 
positive. 
The connection between the $J$ dependence of $\alpha^\mathrm{latt}$ and
the canting angle is highlighted in Fig.~\ref{fig:lnp}b, where the ionic contribution to the
polarization at different values of the magnetic field strength are plotted against
different initial canting angles (obtained by varying $J$).
We observe a linear relationship between the magnetic-field induced
ionic polarization and the zero-field canting angle.
Interestingly, for a critical canting angle ($\theta_c\sim3.2^\circ$), the ionic contribution
to the polarization is zero for all field strengths.
Under these conditions, $\alpha^\mathrm{tot}_{zx}$ is dominated by the electronic
contribution, while for other values of the canting angle the electronic and ionic
components are comparable in magnitude.

To characterize the microscopic origin of the ionic contribution to $\alpha^\mathrm{latt}_{zx}$ in
\lnp{}, we report in
Fig.~\ref{fig:lnpb1u} the phonon mode decomposition of the magnetic-field induced lattice polarization
for different values of $J$.
For \lnp{}, thirteen polar ($B_{1u}$ symmetry) modes enter in determining $\alpha^\mathrm{latt}$. 
As expected, each mode increases in strength linearly with the field (inset of Fig.~\ref{fig:lnpb1u}),
demonstrating that the lattice remains harmonic.
Plotting the contribution to the polarization for each mode for an applied magnetic field of $20$\,T
(Fig.~\ref{fig:lnpb1u}), we find that all modes are important, so that the response is
surprisingly not dominated by the softest lattice modes. 
The zero $\alpha^\mathrm{latt}$ at $\theta_c$ clearly arises from an accidental cancellation
between the positive and negative magnetoelectric responses of the $13$ individual polar lattice modes.

\paragraph*{Conclusion:}
We have demonstrated a new approach for calculating the linear magnetoelectric response of 
materials from first principles. 
The approach, which provides an efficient way to
extract $\alpha$ even for systems with low symmetry, involves
self-consistent application of a magnetic field which acts on the spin degrees of
freedom only in the Zeeman sense. 
Orbital magnetic contributions to responses are neglected within this approach, but both electronic 
and ionic contributions to magnetic-field induced electric polarization are present.

We have demonstrated that the electronic contribution to magnetoelectric responses is not negligible, and
can in fact dominate the total response, in contrast with prior expectations.
Furthermore, we have shown for the case of \lnp{} that the magnetoelectric response is not
dominated by the softest polar lattice modes, and that a partial cancellation of the contributions
from different modes occurs, weakening the magnetoelectric response of this material.
These features suggest that the best route to engineering strong magnetoelectrics may not lie with
strain engineering to induce lattice destabilization, but rather with strengthening the
spin-lattice coupling, for example using strong non-relativistic mechanisms\cite{Delaney/Mostovoy/Spaldin:2009}.

\paragraph*{Acknowledgments:}
We are grateful for fruitful discussions with C.~Ederer and D.~Vanderbilt.
This work was supported by the National Science Foundation under Award No.~DMR-0940420
and by the Department of Energy SciDAC DE-FC02-06ER25794. 
We made use of computing facilities of TeraGrid at
the National Center for Supercomputer Applications and of the California Nanosystems Institute
with facilities provided by NSF grant No.~CHE-0321368 and Hewlett-Packard.
EB also acknowledges FRS-FNRS Belgium.

% Include relevant bibtex files.
%

\end{document}